\documentclass[sigconf]{acmart}

\usepackage{threeparttable}
\usepackage{subcaption}
\usepackage{todonotes}
\usepackage{xcolor}

\usepackage{color, colortbl}
\newcommand{\changed}{} 

\AtBeginDocument{%
  \providecommand\BibTeX{{%
    \normalfont B\kern-0.5em{\scshape i\kern-0.25em b}\kern-0.8em\TeX}}}

\setcopyright{acmcopyright}
\copyrightyear{2021}
\acmYear{2021}
\acmDOI{10.1145/1122445.1122456}

\acmConference[CHI PLAY '21] {2021 Annual Symposium on Computer-Human Interaction in Play}{October 18-21, 2021}{Virtual Event, Austria}
\acmBooktitle{2021 Annual Symposium on Computer-Human Interaction in Play (CHI PLAY '21), October 18-21, 2021, Virtual Event, Austria}
\acmPrice{15.00}
\acmISBN{978-1-4503-XXXX-X/21/10}
\acmDOI{10.1145/XXXXXX.XXXXXX}

\begin{document}

\title{From Flow to Fuse: A Cognitive Perspective}
\author{Kyros Jalife}
\email{kyros.j@northeastern.edu}
\affiliation{%
  \institution{Northeastern University}
  \streetaddress{P.O. Box 1212}
  \city{Boston}
  \state{MA}
}
\author{Casper Harteveld}
\email{c.harteveld@northeastern.edu}
\affiliation{%
  \institution{Northeastern University}
  \streetaddress{P.O. Box 1212}
  \city{Boston}
  \state{MA}
}
\author{Christoffer Holmgård}
\email{christoffer@modl.ai}
\affiliation{%
  \institution{modl.ai}
  \streetaddress{P.O. Box 1212}
  \city{Copenhagen}
  \state{Denmark}
}

\renewcommand{\shortauthors}{Jalife, Harteveld, and Holmgård}

\begin{abstract}
The concept of \textit{flow} is used extensively in HCI, video games, and many other fields, but its prevalent definition is conceptually vague and alternative interpretations have contributed to ambiguity in the literature. To address this, we use cognitive science theory to expose inconsistencies in flow's prevalent definition, and introduce \textit{fuse}, a concept related to flow but consistent with cognitive science, and defined as the ``fusion of activity-related sensory stimuli and awareness''. Based on this definition, we develop a preliminary model that hypothesizes fuse's underlying cognitive \changed{processes}. To illustrate the model's practical value, we derive a set of design heuristics that we exemplify in the context of video games. Together, the fuse definition, model and design heuristics form our theoretical framework, and are a product of rethinking flow from a cognitive perspective with the purpose of improving conceptual clarity and theoretical robustness in the literature.
\end{abstract}

\ccsdesc[500]{Human-centered computing~HCI theory, concepts and models}
\ccsdesc[100]{Human-centered computing~Heuristic evaluations}

\keywords{flow, fuse, cognitive science, theoretical model, design heuristics, video games}

\maketitle

\section{Introduction}

This paper is about those exceptional times when there is no intermediary between us and the sensory stimuli: When we are released from the burden of abstraction and are taken back to a primal state with no thoughts and no meaning. Most of us have experienced this at some point: When dancing, practicing sports, meditating, or playing a video game, we sometimes forget all our concerns and desires, we act without thinking, and become one with the sensory stimuli. Transcendental experiences like this involve so-called altered states of consciousness, such as those described by Maslow's~\cite{maslow1964religions} \textit{peak experiences} and Csikszentmihalyi's~\cite{Csikszentmihalyi1975} widespread concept of \textit{flow}.

Closely tied with the Positive Psychology movement~\cite{Csikszentmihalyi2014}, flow aimed to influence the way both leisure and work experiences are understood and designed for since its beginnings~\cite{Csikszentmihalyi1987}. Books for the general public~\cite{Csikszentmihalyi1990,Kotler2014,Kotler2017} and scientific community~\cite{Engeser2012} were written, an international symposium was recently held~\cite{MaxPlanckInstituteforEmpiricalAesthetics2019}, and even prime ministers and top CEOs have tried to implement Csikszentmihalyi's ideas~\cite{Engeser2012}. Moreover, flow has become a very important concept within the domain of human-computer interaction (HCI) in general~\cite{peifer2020fostering,Finneran2002} and especially within video games, where most designers consider it a clear sign of enjoyment~\cite{Chen2007}.

For all its impact and valuable contribution, Csikszentmihalyi's flow is conceptually vague~\cite{Finneran2002,Novak2000}. As a result, the concept presents critical inconsistencies and lacks a robust theoretical rationale~\cite{Weber2009} (i.e., a systematic and scientifically grounded explanation). On top of this, several different interpretations of the concept have attempted to address these problems, which have contributed to ambiguity in the literature about what flow refers to. However, due to the powerful legacy of Csikszentmihalyi's work, none of them managed to become the new gold standard and the original interpretation is still prevalent, particularly within the field of video games.

Flow is an experiential phenomenon; a concept used to describe a state of consciousness~\cite{Csikszentmihalyi1990}. Because consciousness is thought to be tightly linked to cognitive functions (see Section~\ref{sec:theory}), in this paper we use theories from cognitive science to scrutinize flow's prevalent definition (see Table~\ref{fig:flow scrutiny}) and address the conceptual vagueness in what is arguably its central constituent, the ``merging of action and awareness'', as well as a series of inconsistencies in flow as a whole. To address these issues, we introduce \textit{fuse}. As a word, ``fuse'' means to blend or join to form a single entity. As a concept, this work uses it to describe those experiences where we become one with the sensory stimuli, and defines it as the ``fusion of activity-related sensory stimuli and awareness''.

Following our fuse \textit{definition}, we develop a \textit{model} that presents the \changed{requirement and corequisites} for fuse, as well as a series of mechanisms to facilitate their fulfillment. The \changed{\textit{requirement} and the \textit{corequisites} are interrelated hypotheses that describe a scenario with (1) effortless attention focused on activity-related sensory stimuli (i.e., activity-related sensory stimuli have exclusive control of working memory's [WM] storage, while demand on WM’s central executive is minimal), and (2) effortless avoidance of distractors (i.e., activity-related and unrelated thoughts, and activity-unrelated sensory stimuli do not access WM’s storage, while demand on WM’s central executive is minimal).} The \textit{facilitating mechanisms} aim to identify the circumstances under which fuse is more likely to occur. To complete our work, we demonstrate the practical value of our model by deriving from it a set of \changed{suggested} \textit{design heuristics} to encourage fuse that we exemplify through the medium of video games. \changed{Future work can test and expand these heuristics.} 

Our work---see~\cite{Jalife2019} for a prequel---is part of a growing effort to engage the HCI Game community in critical thinking about the concepts we use. In this case, we do it with flow by introducing a new concept that addresses the issues pointed by our scrutiny, and thus avoids adding ambiguity in the literature by adding yet another interpretation of flow. \changed{Note that we do not discard flow or aim to provide evidence against so-called ``flow experiences''. Rather, we are pointing out the issues with this concept and suggest a more refined alternative. Specifically,} fuse, a product of rethinking flow from a cognitive perspective, describes a narrower, more precise set of experiences; makes a clear distinction between its constituents, its conditions, and potentially associated---but not ontologically relevant---experiential phenomena; and is supported by a theoretical rationale based in cognitive science theory. In its current, preliminary state, our fuse theoretical framework (i.e., our definition, model, and design heuristics; see Figure~\ref{fig:framework}) intends to engage the community in the process of improving our literature's conceptual clarity and theoretical robustness, and has the potential to assist in the design and evaluation of experiences, particularly in the context of video games.
\section{Related Work}
In Csikszentmihalyi's words, flow is ``\textit{the state in which people are so involved in an activity that nothing else seems to matter}''~\cite{Csikszentmihalyi1990}. In this section, we review some of the issues caused by flow's conceptual vagueness. First, we expose arguments presented by previous criticism of flow, and their implications for using the concept. Second, we show how others have attempted to make a better use of the concept by either expanding the theory to other contexts or rethinking it. Finally, we delve into how flow is used in the field of video games. The purpose of this section is not to provide a comprehensive review of all efforts, but to briefly outline the extent of existing efforts relevant to the work presented here.

\subsection{Criticism of Flow}\label{sec:criticism}
Csikszentmihalyi developed the theory of flow mainly based on observations and self-reports. According to Weber~\cite{Weber2009}, since these methods do not give access to the complex neurological states governing the experience, flow's definition and most of its operationalizations aimed at measuring flow, rely on easily identifiable elements that are thought to correlate with the flow experience. As a result, relatability is prioritized over conceptual precision and theoretical robustness, and elements such as the ``balance between difficulty and skills'', or the ``sense of control'', are considered constituents of flow despite them not describing an experience, or empirical research~\cite{Quinn2005} arguing that they are antecedents and consequences of flow, respectively.

Influenced by this, conclusions based on measuring methodologies reliant on self-report (see~\cite{Moneta2014} for a review) have contributed to an imprecise theory. One example is the \textit{flow questionnaire} (FQ)~\cite{Csikszentmihalyi1988}, which asks participants to respond to Likert-like scales about quotes that originated in interviews (e.g., ``I get involved'', ``I enjoy the experience and the use of my skills''), but, according to Moneta~\cite{Moneta2014}, its quotes represent different constructs. Another example is the \textit{experience sampling method} (ESM)~\cite{Csikszentmihalyi1987}, which at random times during daily activities asks participants to report on variables associated with the concept and makes special emphasis in the activity's difficulty and skills. This helped cement the widespread idea that high difficulty and skills is a constituent of flow and not an antecedent, as can be seen in Csikszenmihalyi and Lefevre's work~\cite{Csikszentmihalyi1987}.

As the foundation of flow's theory is based on observations and self-reports, it hinders efforts to develop measuring methodologies based on psychometric variables (see~\cite{Peifer2014} for a review), such as heart rate~\cite{Keller2011,DeManzano2010}, respiratory depth~\cite{DeManzano2010}, and salivary cortisol~\cite{Keller2011}. These methods do not have a clear path for operationalization, and, as Finneran and Zhang ~\cite{Finneran2002} and Moneta~\cite{Moneta2014} argue, they are often based in arbitrary interpretations of the theory, and thus produce incongruent results.

In short, the conceptual vagueness in flow's theory is both motivated and magnified by its high dependence on observation and self-report methodologies. However, there have been efforts directed at improving the theory.


\subsection{Theorizing about Flow}
Multiple efforts looked to expand the theory by proposing ways to adapt the concept to the specifics of video games~\cite{Chen2007,Cowley2008,Lemay2007}, websites~\cite{Huang2003,Novak2000}, and other settings~\cite{Ghani1994,wrigley2013experience,swann2018review}. For example, in the context of media consumption, Sherry~\cite{Sherry2004} posed that flow is the result of the message's difficulty being balanced with the ability to interpret it. In HCI, Finneran and Zhang~\cite{Finneran2002,Finneran2003} developed a model that contemplates the role of artifacts (e.g., a word processor program) mediating the interaction of individuals with tasks (e.g., putting together a resume), and suggested that to elicit flow experiences, it is essential to consider not only the design of the task but also of the artifact. By adapting Csikszentmihalyi's flow to specific fields, however, these efforts have contributed to perpetuate its problems.

Among efforts to rethink flow, we focus on those that do it from a cognitive perspective (see~\cite{Harris2017} for a review). Two stand out due to their conceptual precision and theoretical robustness: Weber et. al.~\cite{Weber2009}, who defined flow---in the context of media consumption---as the conscious awareness of positive affect that results from the synchronization of attentional and reward neural networks, and Dietrich~\cite{Dietrich2004}, who described flow as a state where processed information is exclusively represented in the implicit neural system, resulting in reduced prefrontal cortex activity. Despite their value, these accounts seem to be the result of selective interpretations of the concept and thus contribute to ambiguity in the literature. While Weber at al. focused on ``conscious awareness of positive affect'' despite it not being central to flow's prevalent definition (see Table~\ref{fig:flow scrutiny}) and arguments against it being part of the phenomenology~\cite{Novak2000,Rheinberg2018}, Dietrich did not delve into attentional focus, a key element in virtually all interpretations of flow.

The reviewed works theorizing about flow focused on either adapting the concept to other fields, or on questioning its definition and theoretical rationale. We are part of this second group because we attempt to critically examine the basic assumptions of flow from a cognitive perspective. However, instead of providing a redefinition---and thus contribute to more ambiguity---we address the issues with flow by introducing a new concept that we demonstrate in the context of video games.

\subsection{Flow and Video Games}
Flow is one of the most central concepts in the video game literature. Almost every basic game design book discusses the concept and includes a diagram with challenges and skills as axes and a ``flow channel'' illustrating the optimal experience, which is achieved by balancing the level of challenges with the level of skills. Its popularity has led to several game-specific flow surveys~\cite{kiili2008foundation,fu2009egameflow}. Even if the instruments measure other constructs, such as immersion, engagement, motivation, or player experience, they are inspired by flow. For example, the very recent Player Experience Inventory (PXI)~\cite{abeele2020development} includes the items ``The goals of the game were clear to me'' and ``The challenges in the game were at the right level of difficulty for me'' (see Elements 7 and 9 in Table~\ref{fig:flow scrutiny}, respectively).

Flow has been expanded as a model for evaluating player enjoyment~\cite{sweetser2005gameflow} and a framework for educational games~\cite{kiili2014flow}. Among others, its understanding of video games has been improved by relating it to other models, such as the Technology Acceptance Model (TAM)~\cite{hsu2004people} and User-System-Experience (USE) model~\cite{Cowley2008}, and to specific game aspects or phenomena, such as character identification~\cite{soutter2016relationship} and online game 
addiction~\cite{wan2006psychological}. Efforts have also been made to identify the factors that lead to flow in games~\cite{jin2012toward}. 

Despite its widespread use, there is no clear consensus (and thus understanding) about the concept. In contrast to others \cite{sweetser2005gameflow,Chen2007}, Mekler et. al.~\cite{mekler2014systematic} argue that flow is different from enjoyment. Michailidis et. al.~\cite{michailidis2018flow} argue that flow is not substantially different from the concept of immersion. Scholars also selectively apply the concept for their purposes. For example, in investigating the impact of flow in game-based learning, Hamari et. al.~\cite{hamari2016challenging} operationalized flow as ``heightened challenge and skills''. Therefore, we observe the same challenges with flow in the field of video games as elsewhere, which calls for a careful re-examination or ``rethinking'' of flow.

\section{Theoretical Background}
\label{sec:theory}

In this section, we review the literature used as foundation for our fuse framework, which results from understanding flow as a state of consciousness and scrutinizing it from a cognitive perspective.
\changed{As such, we address work on consciousness, attention and working memory, unconscious processing in action control, self-awareness and mind wandering, and susceptibility to distractions.
All this work contributes to understanding the mechanisms and circumstances that determine what is experienced during fuse (i.e., activity-related sensory stimuli) and what is not experienced (i.e., activity-related and unrelated thoughts, and activity-unrelated sensory stimuli).}

\subsection{\changed{Consciousness, Attention and Working Memory}}
In this subsection, we expose crucial issues surrounding the study of consciousness \changed{and offer a working definition for it; review theory that explains the connection of consciousness with the cognitive functions of attention \changed{and working memory (WM); and provide an account of WM, its capacity, and the selection process for gaining access to it}.} 

\subsubsection{Defining Consciousness}
Consciousness or phenomenal consciousness refers to the existence of a subjective experience. In the words of Nagel~\cite{Nagel1974}, an organism has consciousness ``\textit{if and only if there is something that it is like to be that organism}'' (p.~436). How experience happens at all is the center of the mind-body problem that philosophers have been long debating about~\cite{Mcginn1989}: While some sustain that consciousness is irreducible subjective and, thus, physicochemical processes cannot account for it~\cite{Chalmers1995,Levine1983}, others deny this by taking a materialistic stance~\cite{Dennett1991}. 

Solving this problem is essential for a complete understanding of any experiential phenomena, such as flow or fuse. As of today, however, science can only attempt to explain the mind and consciousness in materialistic terms.
Taking this into account, we consider that someone is conscious---or has awareness, we treat them as synonyms---of something (e.g., a stimulus, an action, an internal state) if there is an experience associated with the processing of information stemming from that something.

\subsubsection{Relating Consciousness to Cognitive Phenomena}
We proceed to consider two rival theories that attempt to explain the mechanisms by which consciousness operates. One of these is the Global Workspace (GW) theory~\cite{Baars1997}, which states that the content of consciousness (i.e., what is experienced) is information produced by specialized unconscious processors that, by means of the \textit{spotlight of attention}, is posted on a network widely accessible by the entire nervous system (i.e., the GW). This theory argues for a distributed model of consciousness. It opposes Crick and Koch's~\cite{Crick1990,Crick2003} theory, in which localized sets of neural events---the neural correlates of consciousness---are responsible for conscious percepts, with attention selecting between competing stimuli. 

For both the distributed and localized models, \textit{attention} acts as the modulator of consciousness~\cite{Baars1997a,Crick2003}. Because a key component of attention (and of our model) is the WM system~\cite{Baddeley2003}, which represents the contents of attention by holding attended information for periods of seconds~\cite{Knudsen2007}, we inquire into its relationship with consciousness. \changed{In this regard, both Baddeley~\cite{Baddeley1993,Baddeley2003} and Cowan~\cite{cowan2008differences} state that consciousness crucially depends on WM (i.e., one is only conscious of information held in WM), while Baars~\cite{Baars2003} argues that WM requires consciousness but not \textit{vice versa}.} Lastly, expressing a middle-ground position, Crick and Koch~\cite{Crick1990,Crick1998} propose that WM is necessary for \textit{vivid} consciousness but that conscious awareness of some primitive features may not rely on it.
As noted, despite there being no consensus regarding the specifics of WM and consciousness' relationship (see~\cite{Persuh2018} for a review), theory suggests a tight link between the two ~\cite{Morsella2013}, which is what we assume in this work.

\subsubsection{\changed{Working Memory, Capacity and Selection Process}}
\label{sec:wm}
\changed{WM is generally conceptualized as a multicomponent system that temporarily stores and processes information. There are several WM models, but the most prominent one is Baddeley’s~\cite{Baddeley2003}, which proposes two modality based temporary stores---the \textit{visuo-spatial sketchpad} and the \textit{phonological loop}---as well as an \textit{episodic buffer} that integrates and holds multimodal information. While these storage components are not present in other models such as Cowan’s~\cite{cowan2008differences}, virtually all WM models agree that the processing of information is conducted by the \textit{central executive}, a component that controls and regulates information through \textit{top-down modulation} processes. These processes, also known as ``executive functions'', most prominently include (1) attention switching between tasks or mental states, (2) updating and monitoring information held in WM to keep only appropriate information in store, and (3) inhibition of dominant, automatic, or prepotent responses when necessary~\cite{miyake2000unity,hofmann2012executive}.}

\changed{The construct of working memory capacity (WMC) is highly dependent on the WM model being used. For example: Cowan et. al.~\cite{cowan2005capacity,cowan2008differences} do not consider temporary stores to be a component of WM, and subsequently argue that WMC solely measures processing tasks from the central executive. However, accounts compatible with Baddeley’s model such as the one of Kane et al.~\cite{kane2004generality} consider that WMC is used both by the central executive and temporary storage components---but mostly by the central executive during tasks with a high cognitive load. In line with empirical evidence~\cite{mayer2007common} (see ~\cite{gazzaley2012top} for a review), we adopt this account. As a result, in our model WMC can be used both by central executive's top-down modulation processes as well as by temporary storage of information associated with sensory-stimuli or with internal states consisting of mental imagery, mental models, mental simulations, metacognitive representations, and memories, that we collectively refer to as ``thoughts''. While sensory stimuli accessing WM reflects an instance of attention being directed ``outwards'', thoughts accessing it reflect an instance of ``inwards'' attention.   
}

\changed{To access and be held in WM’s storage, signals encoding information from sensory stimuli and thoughts need to win a competitive process that selects for the highest signal strength. According to Knudsen~\cite{Knudsen2007}, this can be achieved by (1) high quality of encoded information, (2) bottom-up saliency filters that strengthen signals associated with infrequent or important information, and (3) central executive's top-down modulation processes that update signals of information already in WM, inhibit attention on irrelevant information, and trigger behavioral changes (e.g., eye movements) to improve the quality of encoded information. Since research shows that (3) is associated with the feeling of effort~\cite{mulert2005evidence}, we refer to selection processes involving minimal central executive activity---and thus demanding less WMC---as \textit{effortless attention} (see~\cite{bruya2010effortless} for a review on the subject).}

\subsection{Unconscious Processing in Action Control}
\changed{The goal of this subsection is to provide an understanding of how actions may be executed without activity-related thoughts accessing consciousness}. Thus, we review a series of theories and evidence that account for instances in which information corresponding to intermediate steps leading to action execution is processed without conscious input. These instances of 
``automatic action execution'' are key to our fuse framework.

\subsubsection{Dual Process Theories (DPTs)}
DPTs argue for the existence of two systems of cognition: One presenting rapid, parallel, and automatic processes in which only the final product is posted in consciousness (System 1), and another one with slow and sequential thinking processes, constrained by WMC (System 2)~\cite{Evans2003}. While System 1 is present in other animals and can be used to explain the unconscious processing in action control, System 2 is thought to be uniquely human and responsible for abstract reasoning, planning, the subjective experience of agency, and for monitoring impressions, intentions, and feelings generated by System 1~\cite{Evans2003,Kahneman2011}. Although popular, DPTs have been criticized for lacking conceptual clarity and being based on weak evidence~\cite{Keren2009}.

\subsubsection{\changed{Online Systems}}
\changed{Milner and Goodale~\cite{milner2006visual} describe the existence of \textit{online systems} in the brain that process sensory information and execute corresponding actions automatically, without conscious input. According to Morsella~\cite{Morsella2013}, these systems place minimal demands on WM and lack \textit{``covert processes such as memory, cognitive maps, operations on mental representations, and mental simulation''} (p.~2).
Evidence for online systems include simple motor acts such as licking, chewing, swallowing~\cite{Morsella2009}, and button pressing ~\cite{Morsella2013}, as well as ``\textit{highly practiced and ritualized sensorimotor activities that humans love such as climbing, fencing and dancing}''~\cite{Koch2001}.}

\changed{According to Koch and Crick~\cite{Crick2003,Koch2001}, these systems give rise to \textit{zombie modes} that can only be recognized in retrospect (i.e., they lack explicit memory) and that save us the hundreds of milliseconds that consciousness takes to set in, thus allowing for the rapid execution of complex yet routine tasks. As in the states described by the Recognition-Primed Decision (RPD) model~\cite{Klein1993}, the recognition of patterns by highly experienced people trigger an automatic execution of pre-existent actions plans.}


\subsubsection{\changed{Ideomotor Theories}}
\changed{Ideomotor theories propose that actions are cognitively represented by codes of their perceptual consequences~\cite{hommel2009action}. Following this premise, the Theory of Event Coding (TEC)~\cite{hommel2001theory,hommel2009action} argues that the selection and planning of an action (i.e., the intermediate steps leading to action execution) are indistinguishable processes that can be unconsciously triggered by sensory stimuli activating associated perceptual representations. Meanwhile, building on TEC's main assumptions, the Binding and Retrieval in Action Control (BRAC) framework~\cite{frings2020binding} postulates that since stimuli, responses, and perceptual consequences are all bound together into \textit{even-files}, the re-encountering of any of these features leads to an automatic retrieval of all of them. Once again, this means that the sole perception of sensory stimuli can automatically trigger the execution of actions.}


\subsection{\changed{Self-Awareness and Mind Wandering}}
\changed{In this subsection, we focus on instances of inwards attention in which activity-unrelated thoughts access consciousness. We discuss the nature of the ``self'', provide a definition for self-awareness and review some of its associated functions, and provide an account for the phenomenon known as mind wandering.}

\subsubsection{\changed{Self-Awareness}}
\changed{We are a dynamic system without a constant and we only experience the feeling of being a ``self'' when the model we use to represent this system is posted in consciousness. In other words, the self 
is no more than a representational content associated with a particular brain state. This idea is supported both by philosophical arguments~\cite{Harris2014,Metzinger2005} and neuroscience research, as studies consistently suggest that so-called \textit{default mode network} brain regions that are heavily involved in self-referential processes~\cite{Moran2013, Raichle2015} show a decrease in activity when performing sensory-demanding tasks~\cite{Goldberg2006}. In this context, self-awareness---necessary for theory of mind~\cite{Schmitz2004} and for having a sense of separateness from the environment~\cite{Duval2001,Frantz2005}---requires attention to be at least partially directed inwards, towards a ``\textit{metacognitive representation of any mental state with a propositional content such as beliefs, attitudes, desires, and experiences}'' (p.~941)~\cite{Schmitz2004}.}

\subsubsection{\changed{Mind Wandering}}
\changed{Mind wandering is a ubiquitous phenomenon~\cite{Klein1993,Smallwood2006} in which attention switches towards activity-unrelated thoughts (e.g., memories, mental imagery, and metacognitive representations) as a result of alternative goals becoming the object of mental simulations, which are useful to reduce risk~\cite{Cleeremans2002,Dehaene2001} and to solve complex problems requiring long periods of computation~\cite{Binder1999}. Because mind wandering help us plan crucial future behavior, our cognitive system needs to be sensitive to opportunities that facilitate goal completion, which is why mind wandering---and self-awareness---is often automatically triggered by personally salient stimuli~\cite{Gollwitzer1999,Smallwood2006}}.

\subsection{\changed{Susceptibility to Distractions}}
\changed{In this subsection, we briefly discuss the impact that inwards attention and complex cognition have on the susceptibility to distraction from activity-related sensory stimuli.} 

\subsubsection{\changed{Perceptual Decoupling}}
\changed{Since WMC is limited, inwards attention results in perceptual decoupling~\cite{Schooler2011}, a phenomenon where sensory stimuli perception is diminished. This is because WMC allocated to processing activity-related thoughts---needed for non-automatic action execution---or activity-unrelated thoughts---from self-awareness and mind wandering processes---means less capacity for sensory-stimuli processing, which in turn leads to poorer quality of encoded information and thus reduced access of sensory-stimuli representations to WM storage.}

\subsubsection{\changed{Cognitive Load Theory}}
\changed{Tasks requiring complex cognition (i.e., involving mental simulations, mental modelling, and other operations performed during non-automatic action execution) put a high demand on the central executive component of WM and on WMC~\cite{kane2004generality}. Thus, the higher the cognitive load of a task is, the less the central executive will be capable of keeping activity-unrelated sensory stimuli out from WM once they are perceived. In line with this, Cognitive Load Theory~\cite{lavie2004load,forster2009harnessing} proposes that to minimize distractions from irrelevant sensory stimuli and mind wandering, tasks' cognitive load should be low and sensory load should be high (the latter to leave no space for processing irrelevant information). Note that even though both a high cognitive load and a high sensorial load contribute to increasing task complexity, ``\textit{they clearly have opposite effects on distractor interference}'' (p.~351).}

\subsection{\changed{Key Assumptions and Takeaways}}
\label{sec:takeaways}
\changed{In this subsection, we summarize the key assumptions and takeaways from the reviewed theory, which we use to scrutinize flow and introduce the concept of fuse (Section~\ref{sec:redefinition}). Below, we group the assumptions and takeaways according to the structure of our cognitive-based model for fuse (Section~\ref{sec:model}).}

\subsubsection{\changed{Consciousness, Attention and Working Memory Fundamentals}}
{\begin{enumerate}
    \item Consciousness is modulated by attention and is tightly linked to WM (i.e., vivid consciousness is reliant on WM and information unconsciously processed bypasses WM).
    \item \changed{Attention can be directed outwards, towards sensory stimuli, or inwards, towards internal states, or ``thoughts'' (e.g., memories, mental imagery, and models and simulations).}
    \item \changed{WMC can be used by central executive top-down processes and by temporary storage of information associated with sensory-stimuli or thoughts.}
    \item \changed{The selection process that gives access to WM relies on (1) high quality of encoded information, (2) bottom-up saliency filters, and (3) central executive's top-down modulation processes.}
    \item \changed{Effortless attention occurs when the selection process for gaining access to WM involves minimal top-down modulations from the central executive.}
\end{enumerate}
}

\subsubsection{\changed{Activity-Related Thoughts and Automatic Action Execution}}
\begin{enumerate}    
    \item Automatic action execution occurs when information from internal intermediate steps (i.e., activity-related thoughts) leading to action is processed unconsciously, and can be triggered by sensory stimuli.
    \item \changed{Automatic action execution places minimal demands on WMC and lacks: (1) sense of agency; (2) planning, mental simulations, mental modelling, conscious comparison of alternatives; and (3) explicit memory (i.e., it can only be recognized in retrospect).}
    \item Automatic action execution is common in simple tasks but can also exist in complex sensorimotor tasks, if sufficiently practiced.
\end{enumerate}

\subsubsection{\changed{Activity-Unrelated Thoughts: Self-awareness and Mind Wandering}}
\begin{enumerate}
     \item \changed{Self-awareness and mind wandering are associated with activity-unrelated thoughts, caused by instances of inwards attention often triggered by salient stimuli related to personal goals or concerns.}
    \item Self-awareness requires a representational model of the self to be posted in consciousness and contributes to: (1) sense of agency; (2) theory of mind generation; and (3) sense of being separated from the environment.
\end{enumerate}
    
\subsubsection{\changed{Activity-Unrelated Sensory Stimuli and Distractibility}}
\begin{enumerate}
     \item \changed{Due to limited WMC, inwards attention causes perceptual decoupling, thus interfering with (activity-related) sensory stimuli processing.}
     \item \changed{Tasks involving a high cognitive load and a low sensory load increase susceptibility to distractors (from activity-unrelated sensory stimuli and thoughts).}
\end{enumerate}
\section{From Flow to Fuse}
\label{sec:redefinition}
To define flow, most papers refer at least partially to Csikszentmihalyi's~\cite{Csikszentmihalyi2014} definition, expressed in Table~\ref{fig:flow scrutiny} (see \cite{Chen2007,Cowley2008,DeManzano2010,alexandrovsky2019game,spiel2019surrogate} for examples). However, this definition contains inconsistencies and inaccuracies, some of which were already recognized by Csikszentmihalyi himself. The result is a popular yet conceptually vague definition in most of the literature. Hence, in this section we scrutinize Csikszentmihalyi's flow from a cognitive perspective and, based on this, we introduce a more precise concept supported by cognitive science theory that we call fuse.

\begin{table*}[]
\caption{\changed{Elements of flow according to Csikszentmihalyi's prevalent definition, conclusions from scrutinizing them (with location of theoretical support in parentheses), and their roles in our cognitive-based fuse framework.}}
  \centering
  \includegraphics[width=\linewidth]{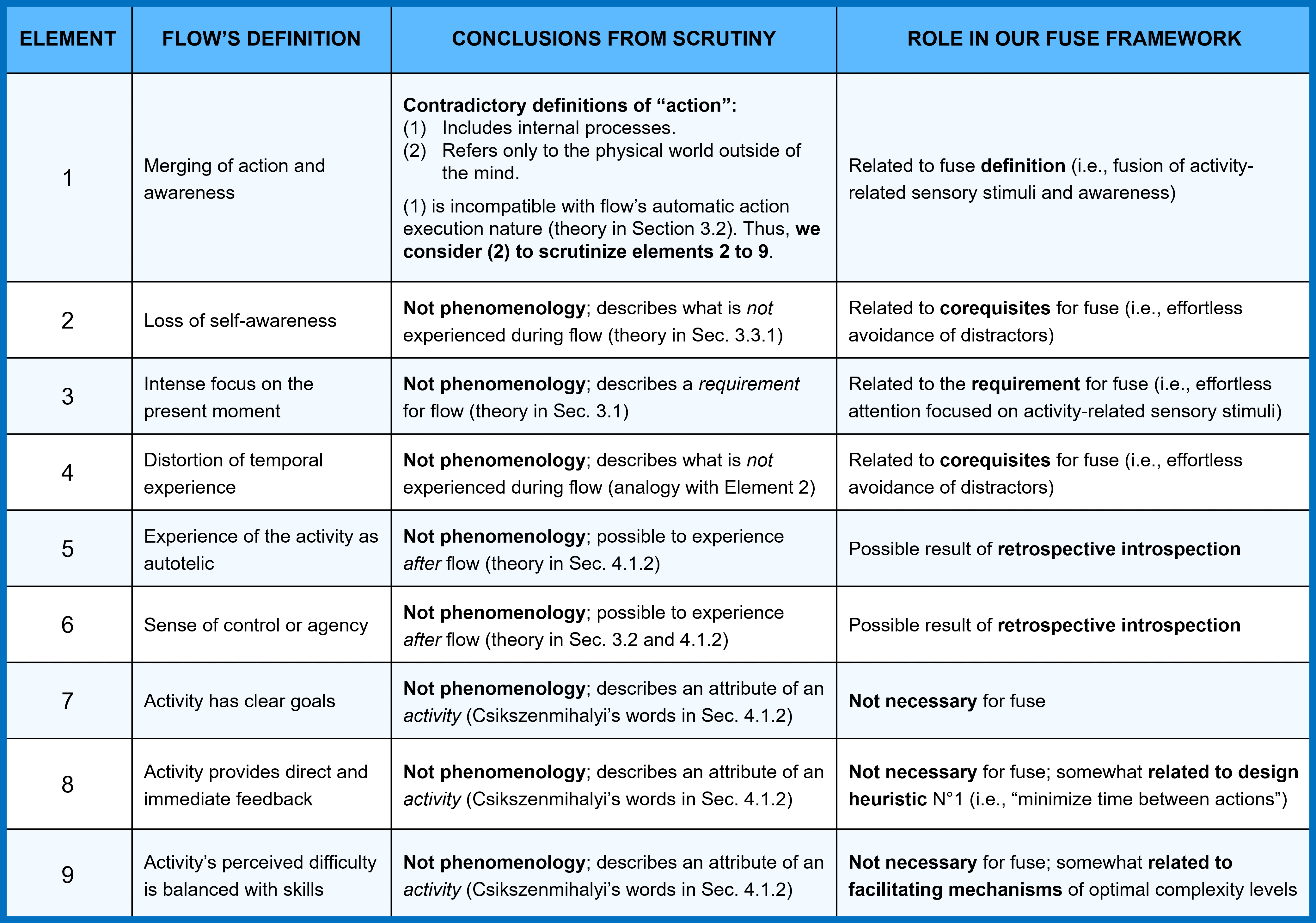}
  \label{fig:flow scrutiny}
\end{table*}

\subsection{A Scrutiny of Flow}
In this subsection, we scrutinize Elements 1 to 9 from Flow's definition, laid out in Table~\ref{fig:flow scrutiny}). Based on reviewed cognitive science theory, we argue that Element 1 contains a critical inconsistency, and that Elements 2 to 9 do not define the experience since it is inaccurate to claim that they are part of flow's phenomenology.

\subsubsection{A Critical Inconsistency}
Csikszentmihalyi observed that ``\textit{the defining feature of flow is intense experiential involvement in moment-to-moment activity}'' (p.~ 230). In other words, the ``merging of action and awareness'' (Element 1). But what is the ``action''? On the one hand, some of Csikszentmihalyi's words seem to indicate that ``action'', or ``moment-to-moment activity'', refers exclusively to the physical world outside of the mind: ``\textit{The merging of action and awareness is made possible by a centering of attention on a limited stimulus field}'' (p.~139), such as when ``\textit{a tennis player pays undivided attention to the ball and the opponent}'' (p.~138), he explained. On the other hand, Csikszentmihalyi also suggests that the ``action'' includes what transpires in the mind, as ``\textit{perhaps in most (flow experiences), one becomes more intensely aware of internal processes}'' (p.~141).

Reviewed theory makes clear that automatic action execution lacks conscious awareness of internal intermediate steps, planning and mental modelling. Thus, Csikszentmihalyi's contradictory \changed{definition of the ``action''} leads to another question: Does flow describe an experience where actions are automatically executed? Again, there is no clear answer. Csikszentmihalyi claims that during flow ``\textit{action follows upon action according to an internal logic which seems to need no conscious intervention}'' (p.~136), as ``\textit{there is little distinction between stimulus and response}'' (p.~137) and ``\textit{one does not stop to evaluate the feedback—action and reaction have become so well practiced as to be automatic}'' (p.~144). These claims are in full conflict with the assertion that during flow people are intensely aware of internal processes, and with the repeated mentions of chess---a game known to produce intense conscious evaluations, mental modelling and planning in both novice and professional players---as an exemplary flow activity.

The conceptual vagueness in what is arguably flow's central constituent (i.e., Element 1) causes irreconcilable inconsistencies in the concept as a whole. \changed{On the one hand, an intense awareness of internal states would mean that actions and reactions cannot be automatic and that there would be a great distinction between stimulus and response. Conversely, if during flow there is only awareness of the physical world outside of the mind, then actions may be executed automatically; however, in that case flow's definition would still contain several inaccuracies, as we demonstrate in the following subsection.
}

\subsubsection{\changed{Flow's Inaccuracies}}
\changed{Even considering the case in which Table~\ref{fig:flow scrutiny}'s Element 1 means that during flow there is only awareness of the physical world outside of the mind and not of internal states, the concept of flow is riddled with inaccuracies. We demonstrate this by showing that Table~\ref{fig:flow scrutiny}'s Elements 2 to 9 do not describe what is experienced during flow (i.e., its phenomenology). Contrary to Element 1, which attempts to describe what \textit{is} experienced (i.e., the ``action''), Elements 2 and 4 describe what is \textit{not} experienced during flow, Element 3 is a requirement for flow, Elements 5 and 6 describe what may be experienced \textit{after} flow, and Elements 7 to 9 refer to attributes of activities hypothesized to induce flow. As a result, we argue that Elements 2 to 9 cannot be part of flow's phenomenology and their inclusion in the definition is a source of conceptual vagueness.}

\changed{\textbf{Elements 2 and 4: What Is Not Experienced.} One can either evoke a self-representational model into consciousness, or not. The ``loss of self-awareness'' (Element 2) refers to the latter, thus pointing to something that is not experienced during flow; a lack of experience, but not an experience of lacking. Similarly, the ``distortion of temporal experience'' (Element 4) refers to the idea that during flow, individuals lose awareness of time passage, which again indicates a lack of experience. Furthermore, note that awareness of time passage being distorted or of a loss of self-awareness would imply that awareness is not merged with action, cancelling Element 1. As a result, Elements 2 and 4 cannot be part of flow's phenomenology.}

\changed{Note that we do not put into question people's reports of losing self-awareness or awareness of time passage during flow: Those happen, but can only be recognized or ``experienced'' in retrospect (Csikszentmihalyi himself made a similar case for Element 6, as we expose later). Although subtle, we believe that this argument is key for improving conceptual clarity in the literature, and that is why our fuse definition and model incorporate it.}

\changed{\textbf{Element 3: A Requirement for the Experience.}}
Since attention acts as a modulator of experience but is not itself experienced, the ``intense focus on the present moment'' (Element 3) is a requirement for flow but not part of its phenomenology. \changed{This was recognized by Csikszentmihalyi when he wrote that “\textit{the phenomenology of flow reflects attentional processes}” (p.~243).}

\changed{\textbf{Elements 5 and 6: After the Experience.}}
We argue that the ``experience of the activity as autotelic'' (Element 5) cannot be part of flow's phenomenology. During flow, extrinsic motivations may not be necessary to engage, but this is a property of the individual's behavior, not of what is experienced~\cite{Engeser2014}. In fact, acknowledging this autotelic trait or even having a conscious awareness of positive affect---the two are sometimes associated---would demand directing attention away from the action \textit{per se}, thus canceling Element 1. Regarding the second, and in line with theoretical~\cite{Novak2000} and empirical~\cite{Rheinberg2018} arguments, we pose that while positive affect is compatible with flow's definition, conscious awareness of it can only be a consequence that results from a retrospective introspective process in which attention is directed towards positive affect after the experience.

Similarly, the ``sense of control or agency'' (Element 6) also cannot be part of flow's phenomenology because it would require an awareness of being in control, which would detract the individual from the action. In terms of the reviewed theory, this is backed by evidence suggesting that the sense of agency is an aspect of conscious action control~\cite{Morsella2013}---or a result of System 2 involvement according to DPTs~\cite{Kahneman2011}---which goes against flow's automatic nature described by Csikszentmihalyi. The sense of agency stems from the perception that one causes their own actions; that there is a correspondence between intention and outcome~\cite{Morsella2013}. As a result, the sense of agency must be joined by a sense of self~\cite{Metzinger2018}. It follows then that the sense of agency (and of ``autotelic activity'') can only be the result of retrospective introspection. Oddly enough, Csikszentmihalyi recognized this, but decided to include Element 6 as part of the phenomenology all the same. In his words:
\begin{quote}
Rather than an active awareness of mastery, it is more a condition of not being worried by the possibility of lack of control. But later, in thinking back on the experience, a person will usually feel that for the duration of the flow episode his skills were adequate to meeting environmental demands (...) (p.~142).
\end{quote}

\changed{\textbf{Elements 7 to 9: Attributes of the Activity.}}
Finally, Elements 7 to 9 do not describe an experiential phenomenon, but attributes of activities regarding their goals, feedback and difficulty. Csikszentmihalyi acknowledged this by calling these elements the ``conditions'' for flow. This is a step in the right direction but introduces another issue: The word ``condition'' implies that without Elements 7 to 9, flow is impossible. Considering a materialistic understanding of the mind, this is fallacious because flow attempts to describe a state of consciousness and as such its only requirement should be the attainment of the brain states responsible for it. Csikszentmihalyi recognized this when he said that a person is potentially capable of experiencing flow regardless of external cues (p.~146). For these reasons, Elements 7 to 9 cannot be part of flow's phenomenology, nor a strict requirement for it.

\subsection{Introducing Fuse}
\changed{Having scrutinized flow's definition, we proceed to introduce a new concept that addresses the critical inconsistency and the inaccuracies that afflict flow: \textit{fuse}. 
}

\subsubsection{Definition}
The fact that flow is used to describe both experiences in which action is automatically executed, and in which there is intense awareness of internal processes, is not consistent with cognitive science theory and creates internal contradictions in the concept's definition. To address this, we take flow's central constituent, the ``merging of action and awareness'', and define ``action'' as the activity-related sensory stimuli. That is, sensory stimuli stemming from (1) the activity's events (e.g., the shape of a tennis ball); and (2) the physical responses triggered by the activity's events (e.g., the muscles contracting when swinging the racket). Following this, we define fuse as the ``fusion of activity-related sensory stimuli and awareness''. According to this definition, during fuse, individuals are aware (i.e., conscious) of their actions but not of internal processes leading to them, thus allowing for an automaticity that breaches the gap between stimulus and response.

\subsubsection{\changed{Key Distinctions with Flow}}
\changed{In contrast to flow, fuse is hereby defined by a single phenomenological element (i.e., the ``fusion of activity-related sensory stimuli and awareness''). While Table~\ref{fig:flow scrutiny}'s Elements 2 to 9 from flow's definition are not part of fuse's definition, most of them are related in some way or another to our concept. The specifics of these relations cannot be fully grasped before developing our model (Section~\ref{sec:model}) and design heuristics (Section~\ref{sec:heuristics}), but we provide an overview of them here.}

\changed{First, the ``intense focus on the present moment'' (Element 3) referring to the attentional processes needed for flow is somewhat related to our requirement for fuse, which states that attention should be effortless and be focused on activity-related sensory stimuli. Second, the ``loss of self-awareness'' and the ``distortion of temporal experience'' (Elements 2 and 4) are related to our corequisites for fuse, which demand effortless avoidance of distractors (such as self-awareness and awareness of time passage) that take attention away from activity-related sensory stimuli. Third, the ``experience of the activity as autotelic'' and the ``sense of control or agency'' (Elements 5 and 6) are a possible result of retrospective introspection, but cannot be experienced during fuse since they require evoking internal states into consciousness.
}

\changed{Lastly, activities with clear goals (Element 7), providing direct and immediate feedback (Element 8), and with a balance between difficulty and skills (Element 9), are not necessary for fuse for the same reasons that Csikszentmihaly acknowledged that they are not necessary for flow. There is, however, a weak relation between Element 7 and our design heuristic of ``minimizing time between actions'', and between Element 8 and our model's facilitating mechanisms describing optimal levels of complexity.
}

\changed{To conclude, compared to flow, fuse is a narrower concept that accounts only for scenarios in which action is automatically executed and there is no awareness of internal states. Unlike flow, it will not be appropriate to describe typical chess experiences, which require conscious processing of the game states and actions. However, it will be for other experiences currently considered ``flow-friendly'', such as the video games aligned with our design heuristics from Section~\ref{sec:heuristics}.
}

\section{A Model for Fuse}
\label{sec:model}
\begin{figure*}[]
  \centering
  \includegraphics[width=\linewidth]{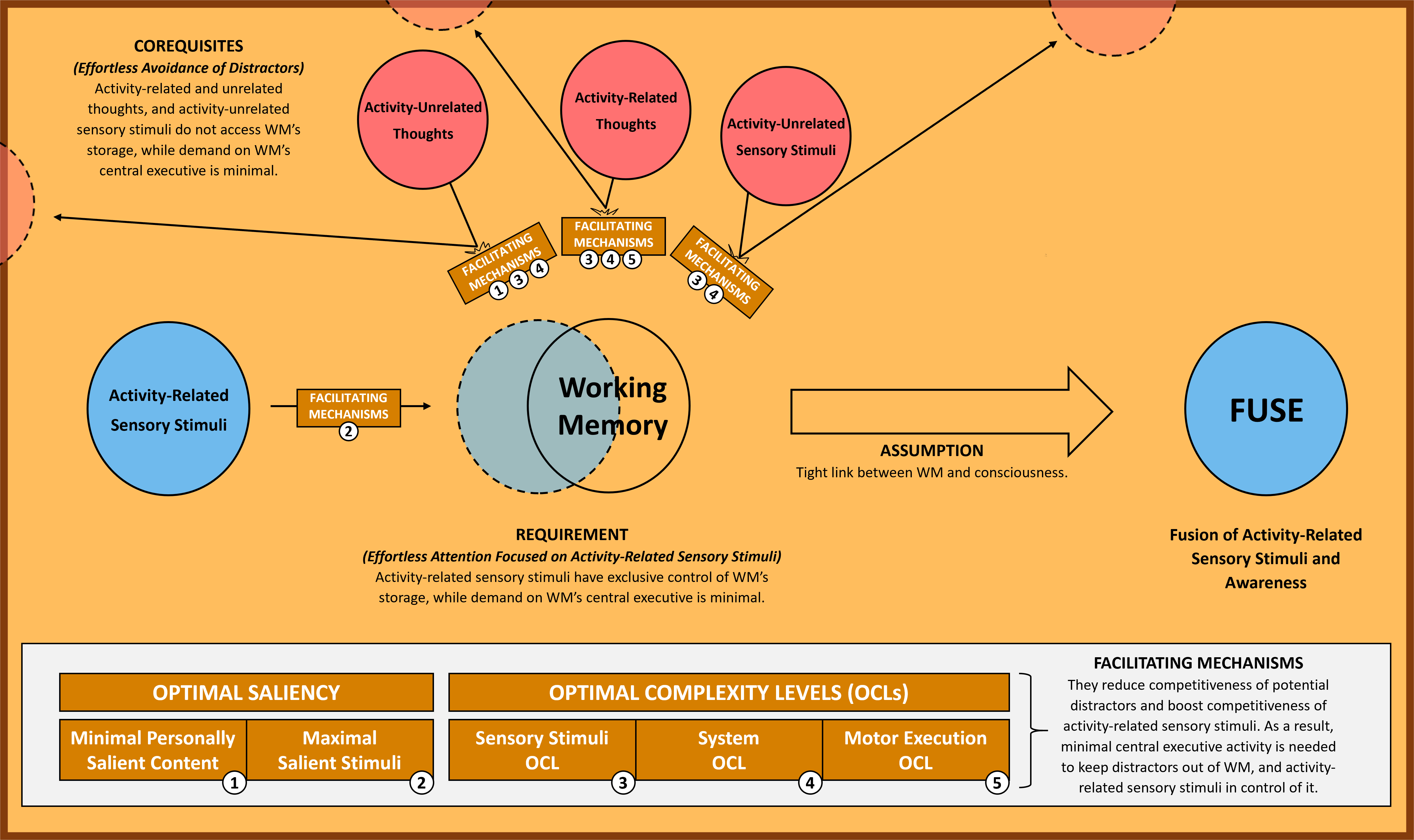}
  \caption{\changed{Facilitating mechanisms are: (1) Minimal Personally Salient Content, (2) Maximal Salient Stimuli, (3) Sensory Stimuli Optimal Complexity Level (OCL), (4) System OCL, and (5) Motor Execution OCL. They contribute to keeping activity-related and unrelated thoughts, and activity-unrelated sensory stimuli out of working memory (WM), and activity-related sensory stimuli in exclusive control of WM, all while producing minimal demands on WM's central executive. As a result, the requirement and corequisites for the fusion of activity-related sensory stimuli and awareness are fulfilled, and fuse occurs.}}
  \label{fig:model}
\end{figure*}

Our fuse definition aims to improve conceptual clarity in the literature and provide a solid base for the development of a robust theoretical rationale. For this rationale, in this section we present the \changed{\textit{requirement} and \textit{corequisites}} for fuse. These are hypotheses about the core cognitive processes underlying the state's phenomenology, corresponding to consciousness being exclusively filled with activity-related sensory stimuli. After this, we depict a series of mechanisms that can be leveraged to aid in the requirement and corequisites's fulfillment, \changed{and explain how this happens by referring to our key theoretical assumptions and takeaways from Section~\ref{sec:takeaways}}. Together, the requirement, corequisites, and the mechanisms form our preliminary model for fuse (see Figures~\ref{fig:model} and ~\ref{fig:OCLs}).

\subsection{\changed{Requirement and Corequisites}}
\subsubsection{\changed{Requirement: Effortless Attention Focused on Activity-Related Sensory Stimuli}}
\changed{Fuse requires WM's storage components to be exclusively controlled by information corresponding to activity-related sensory stimuli (i..e, stemming from the activity's events and triggered responses), and a minimal demand on WM's central executive component, responsible for top-down modulations underlying effortful attention control. This way, a high level of WMC is allocated to the storage and processing of activity-related sensory stimuli, thus avoiding perceptual decoupling and making attention effortless.
}

\subsubsection{\changed{Corequisites: Effortless Avoidance of Distractors}}
\changed{Fuse has three corequisites that occur together with the requirement: (1) activity-related thoughts, (2) activity-unrelated thoughts, and (3) activity-unrelated sensory stimuli do not access WM’s storage components. Like with the requirement, demand on WM’s central executive is minimal, making the avoidance of distractors effortless.
}


\changed{Assuming a tight link between WM and consciousness, the fulfillment of the requirement means that only activity-related sensory stimuli access consciousness, while the fulfillment of the corequisites means that everything else---consisting of activity-unrelated sensory stimuli and of activity-related or unrelated thoughts, such as those associated with mental imagery, mental models, mental simulations, and memories---do not access consciousness. The same way we can define an object by pointing to everything that the object is not, the three corequisites provide an alternative definition for the requirement. Thus, the corequisites are not needed first or are a result of the requirement. Rather, the fulfillment of the requirement equates to the fulfillment of the corequisites, and \textit{vice versa}.}

\changed{The reason our model includes both the requirement and the corequisites is that since they describe opposite ``sides'' of the cognitive processes underlying fuse (i.e., they describe what information does and does not gain access to WM), they are useful for hypothesizing different mechanisms to facilitate fuse.
}

\subsection{Facilitating Mechanisms}
\changed{In this subsection, we present a series of mechanisms aimed at facilitating the fulfillment of fuse's requirement and corequisites. To put it in another way, the mechanisms' purpose is to help keep WM's storage exclusively controlled by activity-related sensory stimuli, and deny access to everything else, while minimizing the need for top-down modulations underlying effortful attention control.
}

\changed{To leverage the cognitive processes underlying fuse, our \textit{facilitating mechanisms} are derived from---and grouped around---two cognitive-oriented aspects: \textit{complexity} and \textit{saliency}. Furthermore, facilitating mechanisms are related to different aspects of an activity. In the case of \textit{optimal complexity} mechanisms, they are related to the activity's \textit{sensory stimuli} (i.e. the input to which the action is a response), \textit{system} (i.e., the interconnected network of components and rules that need to be understood to act), and \textit{motor execution} (i.e., the action itself). Meanwhile, \textit{optimal saliency} mechanisms are related to the activity's sensory stimuli, as well as its \textit{thematic content} (i.e., the objects, concepts and representations that serve more than just action-oriented purposes within the activity and can be related to the external world).
}

\subsubsection{Optimal Complexity}
\changed{Automatic action execution to avoid activity-related thoughts and, more generally, an effortless avoidance of all potential distractors are necessary for the fulfillment of the corequisites. Reviewed theory states that automatic action execution can happen if tasks are simple enough or if complex sensorimotor tasks are sufficiently practiced, and that effortless avoidance of irrelevant thoughts and sensory stimuli requires a low cognitive load and high sensory load. To operationalize these learnings in the context of our fuse framework, we developed parameters to evaluate the activity in relation to its complexity of sensory stimuli, system, and motor execution.
}

\begin{figure*}[]
  \centering
  \includegraphics[width=\linewidth]{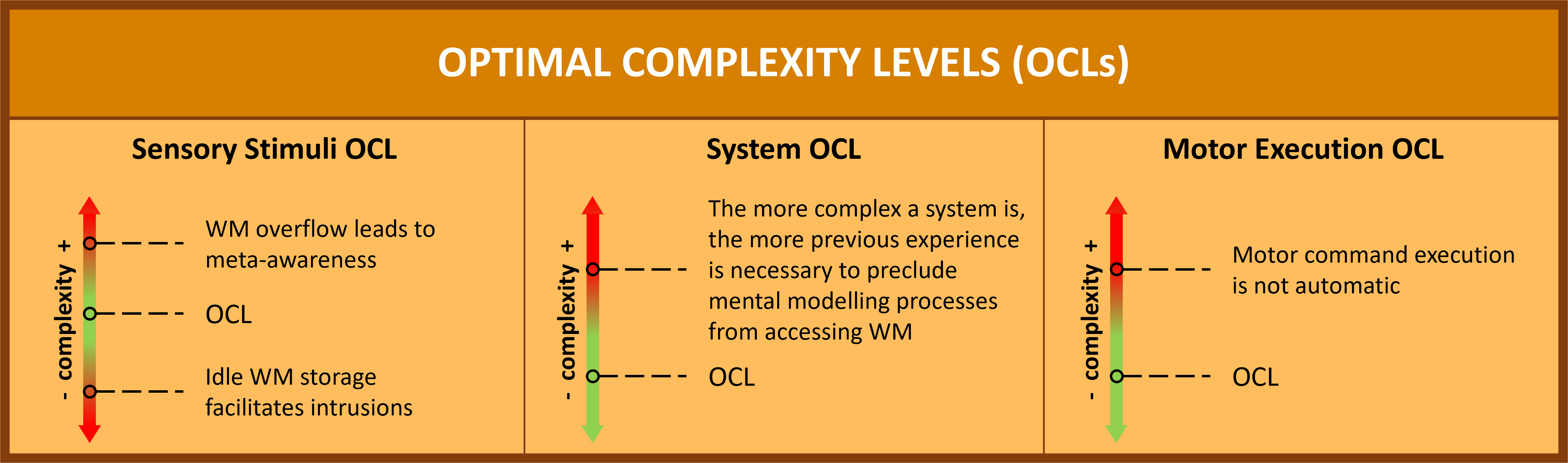}
  \caption{A visual representation of the optimal complexity level (OCL) for each of the three complexity parameters.}
  \label{fig:OCLs}
\end{figure*}

\changed{An \textit{optimal complexity level} (OCL) is the level of complexity in sensory stimuli, system or motor execution such that the individual's skills are balanced with the perceived complexity. OCLs consider not only the real complexity but also the perceived one to take into account scenarios where individuals make an object of attention out of the complexity itself. What follows is a description of the three types of complexities (see Figure~\ref{fig:OCLs} for a visual representation).}


\textbf{Sensory Stimuli Complexity.}
The sensory stimuli complexity challenges skills to effectively and efficiently process sensory information required to successfully perform the activity. 
\changed{In line with Cognitive Load Theory's argument for a high sensory load}, the Sensory Stimuli OCL is sufficiently high, so that WM storage is filled with activity-related sensory stimuli, leaving no space that could be allocated to internal states or activity-unrelated sensory stimuli. However, \changed{we argue that in the context of our fuse framework, this is not enough:} The OCL is also sufficiently low to avoid an overflow of WM that leads to repeated failure, which can itself become an object of attention due to frustration, and which usually comes together with interruptions (e.g., video games' loading screens).

Individuals can potentially compensate for low complexity by increasing their focus, thus remaining fully fused with the activity's stimuli. As an example, think of video games' ``zen modes'' in which sensory stimuli complexity is low due to its irrelevance for achieving goals (it is usually impossible to lose in these modes): Players can be fully attentive of the sensory stimuli but they can also have their attention divided without detriment, which is why an OCL is encouraged. Conversely, sensory stimuli may appear to be too complex, but if individuals know that the activity only requires a scan of superficial layers of information, then WM storage and capacity can suffice. As an example, think of a first-person shooting (FPS) game: There is a myriad of elements that can be attended, but most goals only require recognizing basic shapes and movements to identify targets.

\textbf{System Complexity.}
The system complexity challenges skills to navigate the activity's system. 
This skill is a result of a combination between (1) knowledge of the system and (2) deduction power. The OCL demands minimal or no deduction power \changed{(i.e., a low cognitive load)}, so a plan based on a mental model of the system generated throughout previous experiences (i.e., a pre-existent action plan) is automatically executed. If this was not the case and active mental modeling and simulations were necessary, \changed{then there would be less WMC available to process sensory stimuli---leading to perceptual decoupling---and to keep potential activity-unrelated distractors out of WM}. \changed{In line with the situations described by the Binding and Retrieval in Action Control framework (see Section~3.2.3), a System OCL allows individuals to encounter sensory stimuli patterns that trigger the retrieval of event-files containing appropriate responses, thus avoiding the need to engage in analytical decision-making. 
}

\textbf{Motor Execution Complexity.}
The motor execution complexity challenges skills to effectively and efficiently execute motor commands required to successfully perform the activity. For this type, the OCL is sufficiently low, so that motor commands are executed automatically before selection (only System 1, or online systems, are engaged), bypassing conscious input, and thus producing faster response times.
Potentially, the OCL can increase up to the point where the individual is not physically capable of executing a command. This is the case of professional \textit{NES Tetris}~\cite{Nintendo1989} players that cannot press buttons fast enough to cope with the speed of Level 29, also known as ``the kill screen''.

\subsubsection{\changed{Optimal Saliency}}
\changed{During fuse, activity-related sensory stimuli gain access and retain control of WM storage, and activity-unrelated thoughts---frequent during mind wandering and self-awareness---are denied access to WM. Reviewed theory explains that the selection process towards WM access is influenced by the existence of saliency filters that strengthen signals associated with infrequent or important information (see Section~\ref{sec:wm}), and that self-awareness and mind wandering processes are often triggered by personally salient stimuli. Based on this, we developed the following facilitating mechanisms describing an \textit{optimal saliency} for the activity's sensory stimuli and thematic content. 
}

\textbf{\changed{Maximal Salient Stimuli}}
\changed{Optimal saliency for sensory stimuli occurs when the activity's salient sensory stimulus elements are maximized. By leveraging saliency filters, these elements gain almost instant (25 to 50 ms) access to WM regardless of the nature of the task being performed~\cite{Knudsen2007,Itti2001}, and thus facilitate effortless attention on activity-related sensory stimuli, and the re-catching of attention after a distraction. As an example, think of games with high-contrast elements such as an obstacle displaying a vibrant color or a high-pitched sound breaking silence.
}

\textbf{\changed{Minimal Personally Salient Content}}
\changed{Optimal saliency for thematic content occurs when the activity's personally salient content is minimized. This way, content that could be related to personal goals or concerns are avoided, reducing the risk of triggering self-awareness and mind wandering processes that would cancel fuse. From a representation of a car that reminds someone to go to the mechanic, to a face that resembles an acquaintance, virtually any thematic content can potentially be related to a personal goal or concern. As a result, this facilitating mechanism argues for activities with a high proportion of abstract elements.
}

\subsection{\changed{Fulfillment of Requirement and Corequisites}}
\changed{In this subsection, we refer to our key theoretical assumptions and takeaways from Section~\ref{sec:takeaways} to summarize the facilitating mechanisms' role in encouraging effortless attention and in contributing to each of the corequisites and the requirement.}

\subsubsection{\changed{Effortless Attention and Avoidance of Distractors}}
\changed{Effortless attention occurs when the selection process that gives access to WM involves minimal top-down modulations from the central executive (Section~3.5.1, point (5)). Facilitating mechanisms encourage automatic action execution, discourage self-awareness and mind wandering, reduce susceptibility to distractors, and help to catch and retain attention on activity-related sensory stimuli. As a result, facilitating mechanisms reduce the competitiveness of potential distractors and boost the quality of encoded information from activity-related sensory stimuli, resulting in a minimal need for top-down modulations to keep distractors out of WM, and activity-related sensory stimuli in.}

\subsubsection{\changed{Corequisite: Avoiding Activity-Related Thoughts}}
\changed{To keep activity-related thoughts out of WM, action execution must be automatic (Section~3.5.2, point (2)). For this, tasks must be simple or, if complex, they must be sufficiently practiced (Section~3.5.2, point (3)), which is why the Sensory Stimuli OCL, System OCL, and Motor Execution OCL are ``sufficiently low'' (Section~5.2.1). To comply with Cognitive Load Theory (Section~3.5.4, point (2)), the sensory stimuli OCL is also sufficiently high.}

\subsubsection{\changed{Corequisite: Avoiding Activity-Unrelated Thoughts}}
\changed{To keep activity-unrelated thoughts out of WM, self-awareness and mind wandering must not occur (Section~3.5.3, point (1)). This is why facilitating mechanisms encourage Minimal Personally Salient Content (Section~5.2.2) as well as a low cognitive load and high sensory load (Section~3.5.4, point (2)) aimed at reducing susceptibility to distractions, provided by the Sensory Stimuli OCL and System OCL.}

\subsubsection{\changed{Corequisite: Avoiding Activity-Unrelated Sensory Stimuli}}
\changed{To keep activity-unrelated sensory stimuli out of WM, susceptibility to distractions from irrelevant sensory stimuli must be minimized. Once again, this is why facilitating mechanisms encourage a low cognitive load and high sensory load (Section~3.5.4, point (2)), provided by the Sensory Stimuli OCL and System OCL.}

\subsubsection{\changed{Requirement: Attention on Activity-Related Sensory Stimuli}}
\changed{According to our model, the fulfillment of the three previous corequisites are sufficient for fuse: If activity-related and unrelated thoughts, and activity-unrelated sensory stimuli do not access WM, then only activity-related sensory stimuli controls WM storage. However, we also hypothesize that fuse's requirement can be facilitated through the Maximal Salient Stimuli mechanism, that leverages bottom-up saliency filters to help activity-related sensory stimuli gain and retain access of WM (Section~3.5.1, point (4)). Having said this, we note that our model only considers aspects of an activity from a cognitive perspective, and that there are other perspectives (e.g., narrative theory, user experience design, etc.) that are also important for capturing attention.}




\section{DESIGN  HEURISTICS  FOR  FUSE}
\label{sec:heuristics}
From the model outlined in the previous section, we derive six design heuristics to \changed{encourage} fuse and demonstrate the usefulness of this new concept. \changed{Each of the design heuristics is derived from one or more of the model's facilitating mechanisms, but they are not exhaustive: Our goal was to make sure that all the facilitating mechanisms were covered at least once (see Figure~\ref{fig:framework})}. To illustrate the design heuristics, we exemplify them in the context of video games, an activity that is frequently associated with flow experiences. \changed{We note that since developing the appropriate evaluation instrument is out of the scope of this work, these design heuristics are untested.}

\subsection{No. 1: Minimize Time Between Actions}
Giving players time to make evaluations of the past and projections about the future can lead them to generating mental models and simulations, planning, retrieving memories from previous attempts, and evoking self-related contents as consequence of thoughts about success and failure. As a result, \changed{fuse would not occur}. However, if the player is forced to act intuitively with no pause, \changed{then automatic action execution is more likely to occur and distracting thoughts may not access WM}. As an example, think of an FPS game: Non-discrete choices could lead to endless evaluations of optimal strategies, but because the rhythm is so fast-paced, actions are executed seamlessly.

Because during delays in action production we evoke action-related imagery into consciousness~\cite{Morsella2013}, waiting time before actions needs to be minimal---and many simple actions with little time in-between will be better than a few complex actions with a lot of time in-between. This is why turn-based games are generally not well suited for fuse, except when turns are virtually non-existent. Consider ``bullet'' chess, for example, where each player only has less than 3 minutes in total to plan and execute their moves. While in both this and the standard version of chess players are focused, only with ``bullet'' chess their behavior is aligned with the RPD model where attention is not directed towards internal states and action execution is automatic.

Time between actions must also be reduced to discourage assessments about past events. This is important after failures: The time before play is re-started should tend to be null. To deal with this, games with significant loading times may provide waiting tasks (e.g., a mini game or, in multiplayer games, a spectator camera). \changed{Note that minimizing the time between actions means avoiding periods of inaction, but not that the actual physical movements need to be fast.}

This heuristic \changed{is derived from the System OCL and Motor Execution OCL facilitating mechanisms: Although the heuristic does not modify the game's system and motor execution complexities, it forces players to ignore a portion of them, thus modifying \textit{perceived} complexities. As a result, players can meet their respective OCLs when otherwise complexity levels would be too high}. Notice that this heuristic is related to the traditional flow ``condition'' of direct and immediate feedback (see Element 8 in Table~\ref{fig:flow scrutiny}), although is not quite the same: Its focus is on the players and how they allocate their cognitive resources because the intent is to keep players reacting to the activity regardless of how the activity reacts to them.

\subsection{No. 2: Minimize Relatability}
A less self-relatable game means lower chances of salient stimuli triggering interrupting thoughts. This is achieved by leveraging the game’s audiovisuals, mechanics, and theme/plot. Specifically, audiovisuals should tend to be abstract, mechanics should be sensorimotor-based and theme/plot non-existent or difficult to relate to (see Figure 3 for a classification system visualization). Thus, in line with work suggesting that reflection sits opposite to immersion~\cite{khaled2018questions}, fuse-friendly games discourage players from seeing themselves reflected in the game, so the evocation of personally relevant goals or concerns is less probable.

\begin{figure}[]
  \centering
  \includegraphics[width=1\linewidth]{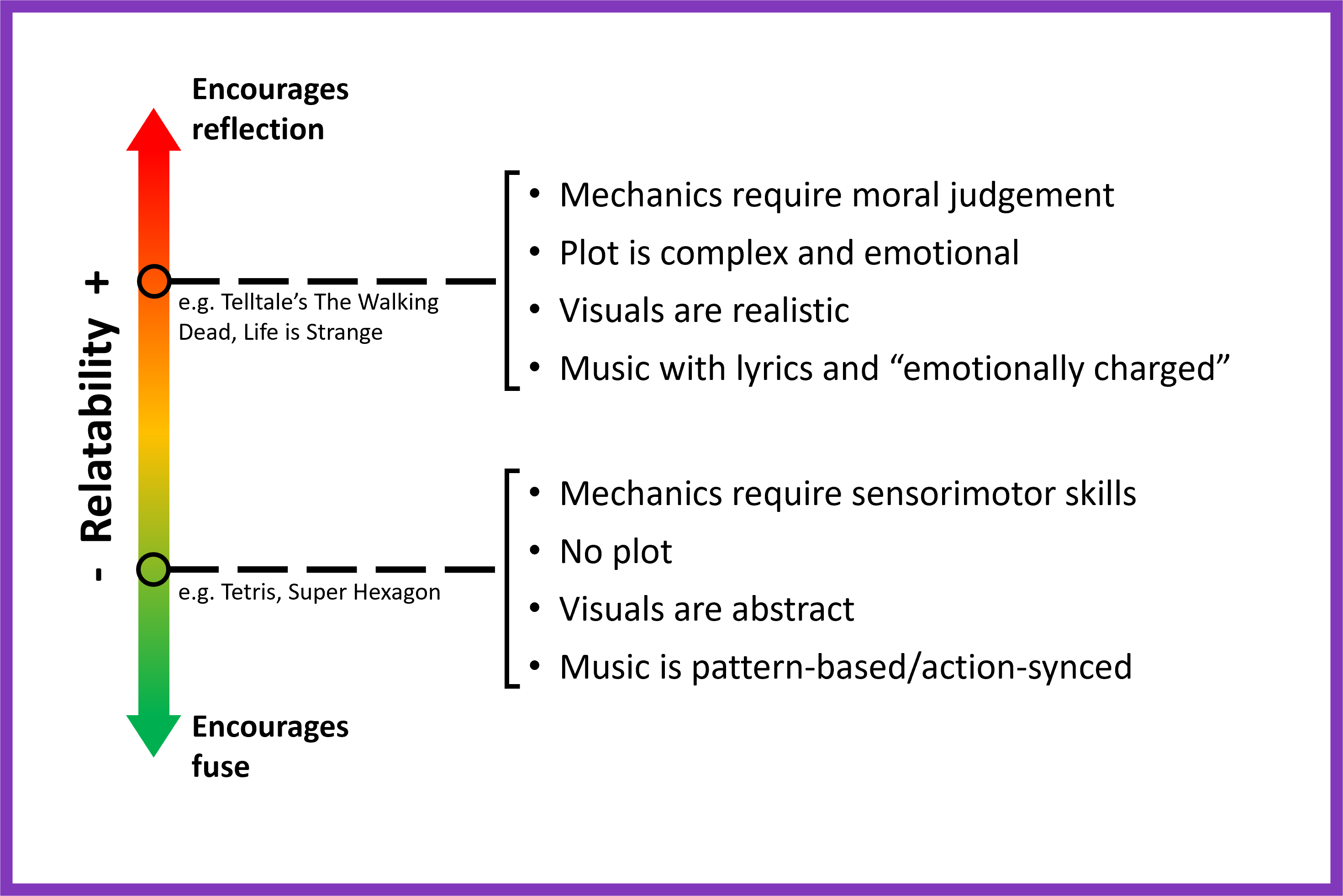}
  \caption{A spectrum to classify video games according to the nature of their mechanics, plot and audiovisuals, which can be either more or less relatable. Fuse-encouraging video games such as \textit{Tetris}~\cite{Nintendo1989} and \textit{Super Hexagon}~\cite{Cavanagh2012} are placed on the green end of the spectrum, while those encouraging reflection such as \textit{Life is Strange}~\cite{DontnodEntertainment2015} and \textit{The Walking Dead}~\cite{TelltaleGames2012} are on the red end.}
  \label{fig:zombie}
\end{figure}

Because fuse and self-awareness are incompatible, activities that intend to elicit fuse must not require self-associated processes such as theory of mind or moral decisions. Thus, video games in which the player interacts with complex characters emulating human emotions or with narratives dealing with human values will not be ``fuse-friendly''. In this context, it is interesting to consider how some game mechanics can help dehumanize even the most hyper-realistic and terrible of scenes---in FPS games players are encouraged to see characters as abstract shapes in motion---and how the insertion of a dehumanizing narrative can transform a game with mechanics as abstract as \textit{Tetris}~\cite{Nintendo1989} in a way that it can be deeply disturbing, as evidenced by the public reaction to \textit{Slave Tetris}~\cite{Thomas2015}.

\changed{This design heuristic is derived from the Minimal Personally Salient Content facilitating mechanism. Thus, in the extreme, it suggests that} to reduce chances of interrupting fuse to a minimum, visual objects should have no apparent real-life utility and extrinsic motivators (i.e., rewards and punishments that are not part of the core mechanics) should be avoided so that mental simulations of actions and evocation of goals are less likely. Lastly, note that although this design heuristic goes against reflection---a significant activity for many players~\cite{mekler2018game}---it facilitates access to a special kind of transcendental state in which players, free from the ``self'', experience deep connection with their senses and the environment’s raw stimuli.

\subsection{No. 3: Leverage Sound Design}
\changed{To avoid over-complicating our model, we did not disaggregate the term ``sensory stimuli'' into ``visual stimuli'', ``auditory stimuli'', and so on. However, given that WM is theorized to have two modality based temporary stores known as the visuo-spatial sketchpad and the phonological loop (see Section~\ref{sec:wm}), image and sound-related signals do not compete for access to WM storage. Thus, since video games are mainly an audiovisual medium, but most of the examples we provided so far are skewed towards visual stimuli, this design heuristic focuses on sound stimuli.}  

In line with research showing that music presence in a game facilitates real-world dissociation~\cite{rogers2019effects}, players should listen to \textit{something} most of the time as this contributes to precluding activity-unrelated sounds from either accessing WM or from being perceived altogether. Even when audio is not part of the game's challenge, as with visuals it needs to meet the Sensory Stimuli OCL: Not too high so the complexity does not become an object of attention and not too low to facilitate the filling of WM (phonological loop) storage. \changed{Furthermore, in line with the Maximal Salient Stimuli facilitating mechanism, audio design should include conspicuous elements that leverage saliency maps to catch and retain attention}. 

To avoid evoking activity-unrelated thoughts, \changed{audio should also comply with the Minimal Personally Salient Content facilitating mechanism}, which means discouraging dialogue and lyrics (see heuristic No. 2). Moreover, to help attention being kept on activity-related sensory stimuli and to facilitate automatic action execution, audio patterns should be synchronized with visual and action execution patterns. The hypothesis behind this is that \changed{conscious evaluation processes are more likely to interrupt effortless attention when patterns are more difficult to recognize (see heuristic No. 6) and when different patterns are not synced, handicapping crossmodal facilitation~\cite{Driver1998}}. Thus, fuse-friendly games should exhibit relatively simple and unified sensory and action execution patterns, which may lead to sound presenting a consistent---and possibly slow~\cite{rogers2019effects}---beat and repetitive notes. An example of this are games such as \textit{Super Hexagon}~\cite{Cavanagh2012} or \textit{Crypt of the Necrodancer}~\cite{BraceYourselfGames2015}, where visual or action execution patterns are synced with audio patterns. \changed{As noted, this design heuristic is derived from the Sensory Stimuli OCL, Maximal Salient Stimuli, and Minimal Personally Salient Content facilitating mechanisms.}

\subsection{No. 4: Help Individuals Meet Their OCLs}
Skills vary between players, as well as within players throughout a playing session due to cognitive and/or physical depletion. Hence, to facilitate \changed{fuse}, players must meet their OCL in each of the three domains, at all times. This can be achieved by embedding the game with either (1) multiple levels of complexity---in the form of coexistent layers or in separate sections---that the players can pick from or (2) a dynamic adjustment system~\cite{Hunicke2005} that does the picking for the player.

One way to attain (1) is to offer multiple goals for each section or level of the game: While an easy goal (e.g., ``reach the end of the level'') would entail just a few objects to scan, simple button combinations to press and would not require elaborate mental models, a more complex goal (e.g., ``collect all gems'' or ``finish the level in less than 90 seconds'') would increase complexity levels. Likewise, (2)---and (1) too---can be attained by keeping an immutable goal (e.g., ``win the race'') while modifying the circumstances in which the action takes place (e.g., more obstacles to avoid).

This heuristic \changed{is derived from the three OCL facilitating mechanisms and} has its similarities with the traditional flow ``condition'' of skills and perceived difficulty being balanced (see Element 9 in Table~\ref{fig:flow scrutiny}). However, we think that the OCL concept is more precise and thus more useful for developing measuring techniques and design strategies.

\subsection{No. 5: Optimize Complexity, Not Difficulty}
\begin{figure}[]
  \centering
  \includegraphics[width=1\linewidth]{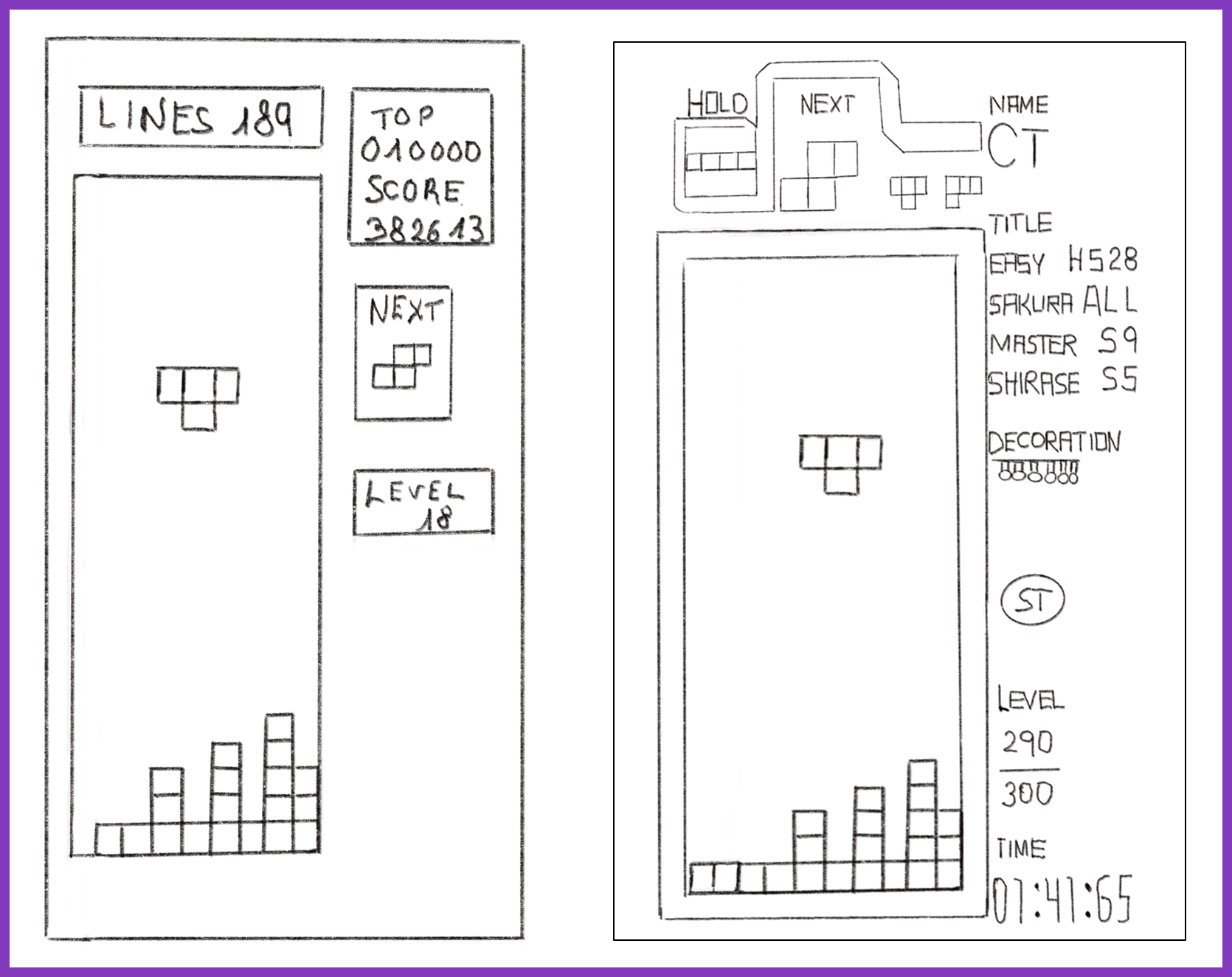}
  \caption{With its added ``hold'' feature and a longer queue in the ``next'' section, \textit{Tetris the Grand Master 3}~\cite{Arika2005} (left) is more complex than \textit{NES Tetris}~\cite{Nintendo1989} (right), but arguably easier since it provides more tools to beat the challenge.}
  \label{fig:tetris}
\end{figure}

Complexity and difficulty sometimes move in the same direction, but they are not the same. When the benefit provided by added features producing extra complexity is higher than the cost of the extra cognitive load, challenges are easier to beat \changed{but meeting the three OCLs requires more skills}. As an example---see Figure~\ref{fig:tetris} for a visualization---think of versions of \textit{Tetris}~\cite{Nintendo1989} that allow players to hold a piece for future use: This feature can be life-saving, but at the same time it adds complexity to the system (and to players' mental model of it), to the execution of motor commands (there is one more button to press), and to the visual interface (the piece on hold is displayed on the screen), which means that more skills are needed for fuse to occur.

Someone could argue that players can just choose to ignore added features like the ones exemplified. While this is a possibility, the sole presence of these features may still contribute to modifying players' perception of the game's complexity levels. In the case of \textit{Tetris}, for example, the presence of an extended ``next'' section queue may trigger distractions (see~\cite{iacovides2015removing} for a related study) related to anxious thoughts about unmet demands, even when the player is not using this information to their advantage.

\changed{This design heuristic---derived from the three OCL facilitating mechanisms---suggests that} a game can also have the exact difficulty level to keep the player engaged, but if the three OCLs are not met then fuse will be less likely to occur. This is what usually occurs with chess, whose system is complex enough to preclude automatic execution from being a constant, even in fully-focused expert players. That said, this type of high-complexity scenario can still be optimized for fuse by implementing heuristic No. 1.

\begin{figure}[]
  \centering
  \includegraphics[width=1 \linewidth]{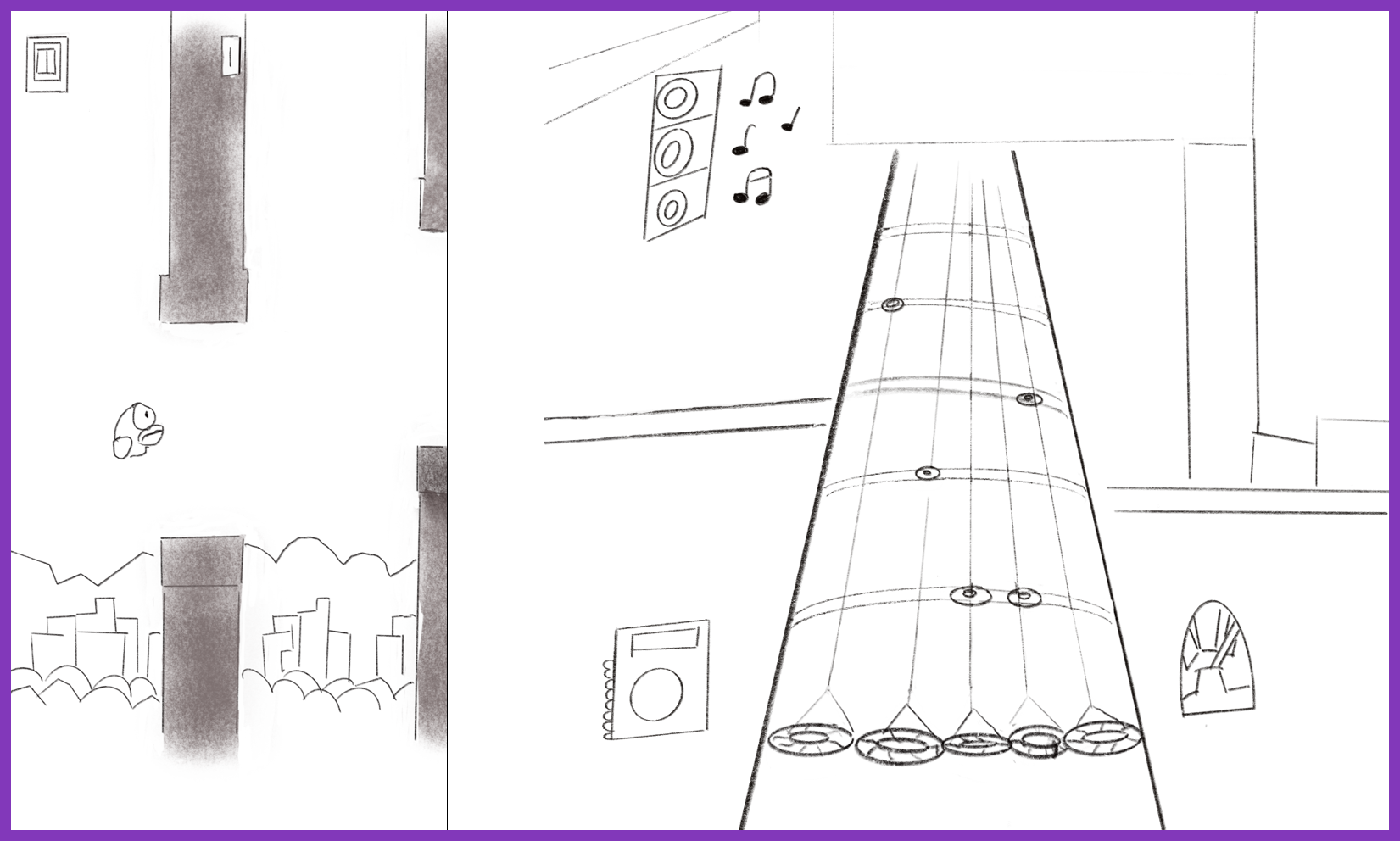}
  \caption{In \textit{Flappy Bird}~\cite{Nguyen2013} (left) choices (i.e., when to make the bird flap) have short-term interdependency: Once an obstacle is averted, previous choices do not matter to predict contingencies. Furthermore, the system’s rules are easily recognizable, making it simple to understand how the transition between states work. \textit{Guitar Hero III}~\cite{NeversoftEntertainment2007} (right) shares these traits: Choices have virtually no interdependency.}
  \label{fig:flappy}
\end{figure}

\subsection{No. 6: Facilitate Learning with Predictable Systems}
\changed{To facilitate automatic action execution, players must meet their System OCL, which means that they must navigate the activity's system based on knowledge---not deduction relying on mental modelling and simulations---of the system's possible states and their respective contingencies}. This entails a learning process that is facilitated when the system is composed of easily recognizable patterns (i.e., it is predictable). However, this does not imply that at any given point a player should be able to predict the game's future states. Rather, it means that they should not be surprised by any of these states because they lie within learned parameters clearly defined by the system's rules.

Some video games, especially narrative ones, are designed to be unpredictable: They intend to surprise. The ``problem'' with this is that players are encouraged not only to engage in conscious evaluation processes and reflection~\cite{whitby2019one}, but also to make an object of attention out of the surprise itself. Unexpected events push us to revise and update a narrative of our own that we use to explain reality (i.e., our mental models). Contrary to this, games with predictable systems do not defy expectations: After building the corresponding mental model we can play non-stop without consulting with ourselves about the events that transpire.

The less contingencies are involved, the easier it is to learn the system and build a mental model of it that allows for automatic action execution. To build simple enough systems, choices should have low and/or short-term interdependency (examples in Figure~\ref{fig:flappy}). This, plus simple enough rules make for low system complexity and thus allow players to meet their System OCL more easily.

This heuristic\changed{, derived from the System OCL facilitating mechanism, }is important because high system complexity may be one of the main obstructions for fuse. To illustrate this, consider once more the case of \textit{Tetris} where expert players know the system so well that they can play until task complexity reaches their sensorimotor skills limit (tapping speed, mostly). This does not happen in many games, as we hypothesize that the limiting variable for fuse is usually decision making and not physical capacity.

A traditional flow condition is for the activity to have clear goals (see Element 7 in Table~\ref{fig:flow scrutiny}). According to our model for fuse, a system should have simple enough rules producing patterns that the individual can recognize and react to, irrespective of the existence of clear goals. In this regard, think of a skillful guitar player experiencing fuse during an improvisation solo: While there are no clear goals, patterns formed by preceding notes are clear and cause the automatic execution of succeeding ones.
\section{Theoretical Framework}
Despite its flaws, the concept of flow as theorized by Csikszentmihalyi is still prevalent. To contribute to conceptual clarity and theoretical robustness in the literature, we conducted a cognitive-based scrutiny of flow. \changed{We concluded that flows' prevalent definition is not consistent with cognitive science theory on automatic action execution and contains several inaccuracies in the description of its phenomenology. To address this, we introduced fuse. In contrast to flow, fuse is defined by a single phenomenological element: the ``fusion of activity-related sensory stimuli and awareness''. As a result, fuse is a narrower concept than flow: It only accounts for scenarios where actions are automatically executed and there is no awareness of internal states, or ``thoughts''.}

\changed{Our next step was to hypothesize about the cognitive processes underlying fuse. This led to our model's (1) requirement and (2) corequisites, which describe a scenario of effortless attention (i.e., minimal demand on working memory's [WM] central executive), while (1) activity-related sensory stimuli have exclusive control of WM’s storage, and (2) activity-related and unrelated thoughts, and activity-unrelated sensory stimuli do not access WM’s storage.}
\changed{We proceeded by using theory on the mechanisms of working memory, unconscious processing in action control, self-awareness and mind wandering, and susceptibility to distractions, to conceive a series of facilitating mechanisms that may contribute to the fulfillment of the requirement and corequisites. In this regard, we propose that an activity should present (1) Minimal Personally Salient Content and (2) Maximal Salient Stimuli, while the perceived complexity of (3) Sensory Stimuli, (4) System, and (5) Motor Execution should be balanced with the individual's skills (i.e., the activity should present optimal complexity levels, or OCLs).}

\begin{figure*}[]
  \centering
  \includegraphics[width=\linewidth]{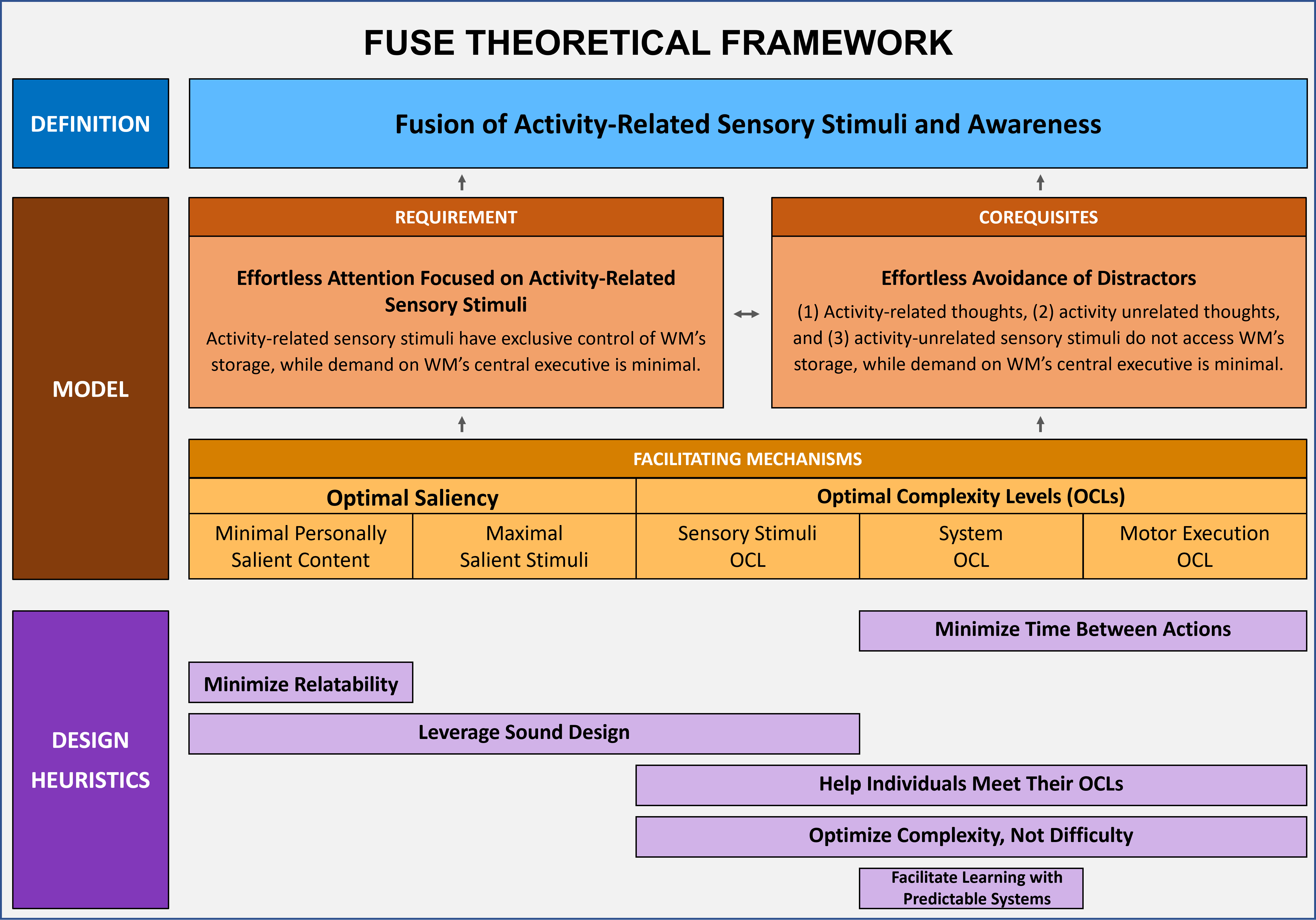}
  \caption{Our fuse theoretical framework, comprised of the fuse definition (top), the model with the cognitive requirement and corequisites for fuse and mechanisms that facilitate their fulfillment (middle), and \changed{design heuristics derived from the facilitating mechanisms} (bottom).}
  \label{fig:framework}
\end{figure*}

\changed{Finally, from the facilitating mechanism we derived six design heuristics to encourage fuse, and exemplified them in the context of games}: (1) Minimize Time Between Actions (2) Minimize Relatability (3) Leverage Sound Design (4) Help Individuals Meet Their OCLs (5) Optimize Complexity, Not Difficulty and (6) Facilitate Learning with Predictable Systems. Figure~\ref{fig:framework} shows our fuse theoretical framework, comprised of the definition, model, and design heuristics.

\section{Discussion}
\label{sec:discussion}
In this section we answer key questions about our work's motivation, approach, contribution, implications, and future steps.

\subsection{Do We Really Need to Revisit Flow?}
Csikszentmihalyi's flow is prevalent in the video game literature. For example, by looking at works published in CHI PLAY's proceedings during the last three years, we found that out of 121 research articles, 28 (23\%) used the concept of flow, and that out of these, 16 (57\%) did it by directly citing Csikszentmihalyi's work, 9 (32\%) by citing survey methods based on Csikszentmihalyi, like the Game Experience Questionnaire~\cite{ijsselsteijn2013game} or GameFlow~\cite{sweetser2005gameflow}, and the remaining 3 (11\%) did it without providing citations but were clearly referring to flow as conceptualized by Csikszentmihalyi. \changed{Furthermore, at the recent CHI'21 conference, several works discuss flow, for example, recommending to ``Use flow to structure the learning curve of human-AI interaction''~\cite{zhu2021player}, and using the concept to suggest that ``flow encourages task focus''~\cite{david2021flow}.} 

In the vein of recent works that aimed to engage the community in critical thinking about other popular concepts and theories, such as Tyack and Mekler~\cite{tyack2020masterclass,tyack2020self} did with Self-Determination Theory, we believe that there should be an open discussion about flow as theorized by Csikszentmihalyi. The difference between these works and ours, is that instead of engaging the community through a literature review, we did it by scrutinizing the theory from a cognitive science perspective.

\subsection{Why Use a Cognitive Perspective?}
Flow is meant to describe a special state of consciousness, which, according to reviewed theory, is modulated by the cognitive function of attention. Because of this, cognitive science is useful to expose some of flow's issues \textit{and} to provide theoretical support to any alternative concept attempting to describe a state of consciousness, like fuse does. Furthermore, our cognitive approach intends to promote discussion about the mechanisms underlying transcendental experiences that create altered states of consciousness, such as those described by Maslow's~\cite{maslow1964religions} peak experiences and Csikszentmihalyi's~\cite{Csikszentmihalyi1975} flow.
The resulting understanding will contribute to the design of experiences and the development of much-sought methods of objective measurement.
While we argued for the logic of rethinking flow from a cognitive perspective, we acknowledge that such a complex concept can be scrutinized from a wide range of disciplines and perspectives, and welcome the community to further scrutinize flow. \changed{We also acknowledge that playing games is not just about these ``flow experiences'' and that ``ordinary player experiences''~\cite{tyack2021off} should also be of consideration. Perhaps a different concept than fuse will help explain such experiences.}   

\subsection{Why Introduce a New Concept?}
Besides the conceptual vagueness and ambiguity that affects flow, the HCI literature, particularly within video games, is riddled with confusion surrounding terms like immersion, engagement, motivation, enjoyment, presence, and flow, as oftentimes these terms are used interchangeably. While reducing the amount of concepts we use would solve this problem, it would also reduce our capacity to understand experiences and to design for them. Instead, the solution lies in improving conceptual clarity and reducing ambiguity in the literature. Fuse, which is a product of rethinking flow from a cognitive perspective, is a step in this direction.

As demonstrated by our design heuristics, the introduction of fuse may also contribute to the development of new design approaches, particularly for video games. In this regard, our six design heuristics for fuse are either novel (e.g., heuristic No. 2: ``Minimize Relatability'') or represent an alternative to some traditional flow ``conditions'' (e.g., heuristic No. 1, ``Minimize Time Between Actions'', is related to flow's ``direct and immediate feedback'' condition).

\subsection{How Exactly is Fuse Different from Flow?}
Unlike flow, fuse does not contemplate experiences in which there is awareness of internal processes. This means that traditional flow experiences in which attention is directed inwards for planning, memory retrieval, mental simulations, and conscious comparison of alternatives, do not fit our concept. Typical chess experiences, for example, can be described with flow but not with fuse. Since fuse focuses exclusively on experiences where action execution is automatic, research on unconscious processing in action control suggests that ``fuse-friendly'' activities are those characterized by simple sensorimotor actions (e.g., playing the game Flappy Bird), as well as more complex tasks that can be automated with enough practice (e.g., playing a complex piano piece). Note that these activities could also be classified as flow-friendly since both automatic action execution and intense awareness of internal processes are part of flow's proposed phenomenology, despite theory suggesting that this is contradictory. While less likely to happen, we further note that fuse can potentially be experienced in ``non-fuse-friendly'' activities, which includes the aforementioned chess experience but also playing a narrative-based video game, as long as the pertinent brain states occur (i.e., if attention is not directed inwards or towards activity-unrelated sensory stimuli). 

\subsection{What Does it Mean for Flow?}
Our work aims to engage the community in rethinking flow. Fuse is one possible outcome of this critical examination, and it may be better suited to describe some of the experiences that as of today are associated with flow. Despite this, we believe that flow can still be a useful concept. For example, it can be used to describe experiences in which there is intense awareness of internal processes, or in which deep concentration is joined by positive affect. Furthermore, flow will undoubtedly continue to have an important place in the video game literature, where designing for flow is frequently considered a synonym of not making the game too hard or too easy, to avoid producing states of boredom and anxiety.

\subsection{What is Next for Fuse?}
Our fuse theoretical framework is not definitive. Before we can properly validate it---a process that would take multiple years---we need to engage the community in a constructive debate that will undoubtedly serve to adjust our theory. Only after this our model for fuse should be operationalized to allow for precise predictions and testable hypotheses. \changed{As a result, we limited the current work to exemplifying the practical value of the theoretical model of fuse through design heuristics and considered discussing measures of fuse out of scope. Future work, however, should consider the ways in which fuse can be measured, ranging from biometrics and game data to self-reports and field observations.} 
For instance, a survey could ask video game players if they recall comparing choices or making plans during gameplay, or if they made an effort to decipher subtext. \changed{Regarding the use of surveys, we note here that the criticism of flow (Section~\ref{sec:criticism}) is not about the use of self-reports or other observational methods. Rather, the criticism is that flow has been theorized \textit{from} self-reports and other observational data, and is thus not grounded in existing (cognitive science) theory. Here we suggest developing a survey about fuse, which unlike flow, is grounded in (cognitive science) theory.} 


As for our design heuristics, they were developed with games in mind, to illustrate the concept’s usefulness for the HCI Game community. However, we believe that they can be adaptable to other media. In addition, as mentioned, more design heuristics can be derived from our model. For example, designers could use research on saliency maps to generate detailed guidelines for fuse-encouraging user interfaces, or research on theory of mind to suggest specific patterns of character behavior that do not trigger self-awareness. \changed{Finally, work is needed to test the heuristics as they remain hypothetical at the moment, both in terms of how designers interpret and act on the heuristics in their practice as well as if the heuristics actually achieve fuse.}


\section{Conclusion}
Despite other work attempting to address flow's conceptual vagueness by producing their own interpretations of the concept, Csikszentmihalyi's definition is still prevalent. To engage the HCI Game community in the process of rethinking flow and to avoid adding ambiguity to the literature, we exposed inconsistencies in flow's prevalent definition through a cognitive-based scrutiny, and addressed these by introducing fuse, a new concept meant to describe experiences where consciousness is filled exclusively with activity-related sensory stimuli. Together, our fuse definition, model and design heuristics form our preliminary fuse theoretical framework, which represents a step towards improving conceptual clarity and theoretical robustness in the literature, particularly within the field of video games, where flow takes a prominent place.


\bibliographystyle{ACM-Reference-Format}
\bibliography{main-submission}


\end{document}